\documentclass[reprint,superscriptaddress,amsmath,amssymb,aip,apl]{revtex4-1}

\usepackage[version=3]{mhchem} 
\usepackage{graphicx}
\usepackage{hyperref}
\usepackage{physics}
\usepackage{braket}
\usepackage{siunitx}
\newcommand{\code}[1]{\texttt{\detokenize{#1}}}
\newcommand{\JSC}[0]{$J_\text{SC}$ }
\newcommand{\VOC}[0]{$V_\text{OC}$ }

\begin{document}



\title{Ab initio calculation of the detailed balance limit to the photovoltaic efficiency of single p-n junction kesterite solar cells}

\author{Sunghyun Kim}
\affiliation{Department of Materials, Imperial College London, Exhibition Road, London SW7 2AZ, UK} 
\affiliation{Department of Materials Science and Engineering, Yonsei University, Seoul 03722, Korea}

\author{Aron Walsh}
\affiliation{Department of Materials, Imperial College London, Exhibition Road, London SW7 2AZ, UK} 
\affiliation{Department of Materials Science and Engineering, Yonsei University, Seoul 03722, Korea}
\email{a.walsh@imperial.ac.uk}

\date{\today}

\begin{abstract}
The thermodynamic limit of photovoltaic efficiency for a single-junction solar cell can be readily predicted using the bandgap of the active light absorbing material. Such an approach overlooks the energy loss due to non-radiative electron-hole processes. We propose a practical \textit{ab initio} procedure to determine the maximum efficiency of a thin-film solar cell that takes into account both radiative and non-radiative recombination. The required input includes the frequency-dependent optical absorption coefficient, as well as the capture cross-sections and equilibrium populations of point defects. For kesterite-structured \ce{Cu2ZnSnS4}, the radiative limit is reached for a film thickness of around \SI{2.6}{\micro\meter},  where the efficiency gain due to light absorption is counterbalanced by losses due to the increase in recombination current. 
\end{abstract}


\maketitle

The efficiency of solar cells is determined by the thermodynamics of light and matter.
\cite{}
For single \textit{p-n} junction solar cells,
the first theoretical investigation on the efficiency was given by Shockley and Queisser (SQ),\cite{Shockley:1961co}
where bandgap is the sole parameter that determines the energy loss mechanisms including light absorption, hot carrier cooling, and radiative recombination.
In this model, the absorbance and non-radiative recombination rate are generally assumed to be 1 and 0, respectively.
For this reason, bandgap is often the main descriptor used in  computational materials discovery for solar energy applications.\cite{castelli2012computational}

\begin{figure}[tb]
	\includegraphics[width=8cm]{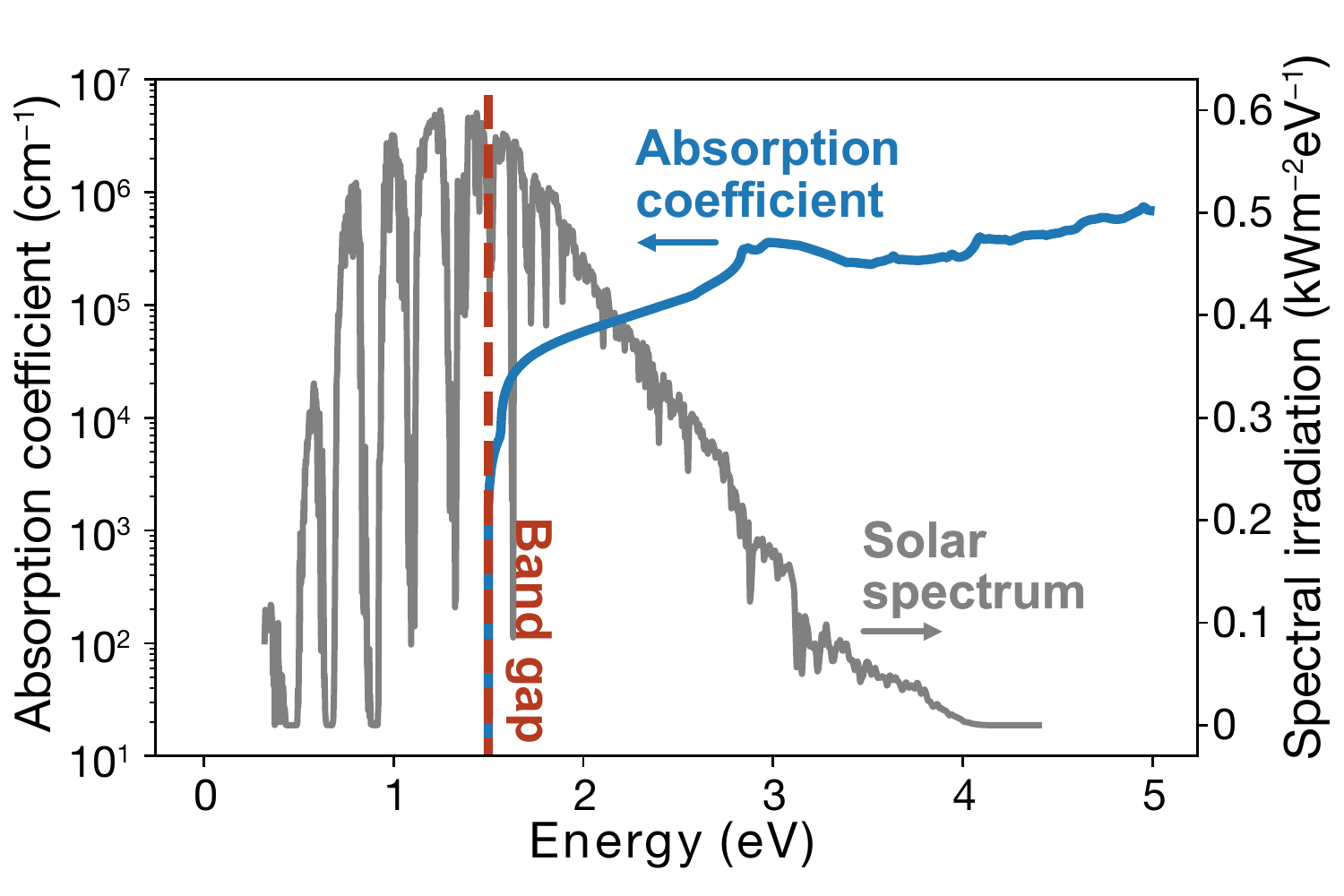}
	\caption{Calculated optical absorption coefficient of \ce{Cu2ZnSnS4} (blue line) from Wannier function interpolation of a hybrid density functional theory band structure.
		The AM1.5g solar spectrum is shown in a grey line.
		The electronic bandgap is indicated by a vertical dashed line.
	}\label{fig:alpha}
\end{figure}

The wavelength-dependent optical absorbance of materials can be calculated from first-principles  (e.g. Fig. \ref{fig:alpha}).
Yu and Zunger \cite{Yu:2012jv} proposed the \textit{spectroscopic limited maximum efficiency} (SLME) that takes into account the finite absorptivity $\alpha$. 
This quantity be calculated using a variety of approaches including density functional theory (DFT)\cite{Hohenberg1964, Kohn1965} and many-body perturbation theory based on the \textit{GW} approximation.\cite{Hedin:1965hi}
The radiative efficiency --- the fraction of radiative recombination over the total recombination current --- is conventionally taken as unity in the SQ model. 
Instead, SLME estimates the fraction of radiative recombination $f_r = \mathrm{e}^{\flatfrac{-\Delta}{k_\mathrm{B}T}}$, 
where $k_\mathrm{B}$ and $T$ are the Boltzmann constant and temperature, respectively.
$\Delta$ is the difference between energies of the absorption edge and the band gap.
The factor $f_r$ is based on the assumption that carriers that cannot recombine radiatively eventually recombine non-radiatively via, for example, an Auger process.
However, the main source of non-radiative recombination is often crystal imperfections,
which trigger trap-mediated non-radiative recombination\cite{Park:2018et} whose steady-state recombination kinetics follows Shockley-Read-Hall (SRH) statistics.\cite{Shockley:1952it,Hall:1952iz}

%

Many classes of materials have been studied for thin-film solar cells.
In particular, \ce{Cu(In{,}Ga)Se2} and \ce{CdTe} have been commercialized.
Most candidate technologies fail to exceed 20\% efficiency even though they exhibit bandgaps within a 1--1.5 eV target window that should support at least 30\% efficiency in the SQ model. 
One example is kesterites, based on quaternary \ce{I_2 \hyphen II \hyphen IV \hyphen VI_4} minerals, which have been studied extensively for thin-film devices.
Despite a lot of effort made to improve kesterite solar cells during the last decade,
the performance struggles to break the 13\% efficiency ceiling.\cite{Son:2019hm, Wang:2013gs}
The biggest challenge is the open-circuit voltage deficit with respect to the SQ limit, which is common in many emerging photovoltaic technologies.\cite{wong2019emerging}
To overcome the limited predictive power of models based on the physical properties of the ideal crystal alone,
one needs to address other origins of non-radiative recombination. 
Recently, we proposed trap-limited-conversion efficiency (TLC),\cite{Kim:2020gj} 
the upper bound of photovoltaic efficiency of a crystal containing an equilibrium population of native defects and their associated carrier capture cross-sections. However, in this prior work complete absorption of photons was assumed above the bandgap.   

In this study, we evaluate the upper bound of photovoltaic efficiency including both finite absorptivity and defect-mediated non-radiative recombination from first-principles.
Kesterite solar cells suffer from significant non-radiative losses due to point defects.\cite{chen2013classification,hadke2019suppressed,giraldo2019progress}
The optical absorption loss is predicted not to be an efficiency-limiting factor when the absorber layer is thicker than \SI{2}{\um}.\cite{Grenet:2018ia} 
We present a workflow for predicting the optimal thickness of absorber layer drawing from \textit{ab initio} calculations.

\textit{Photovoltaic theory:}
The essential metric to evaluate a photovoltaic material is the ability to absorb light.
Since we intended to estimate the upper bound of the efficiency of photovoltaics, the reflectivity at the top surface will be taken as 0.
%
The absorbance $a$ at photon energy $E$ is given by 
\begin{equation}\label{eq:absorptivity}
    a(E;W) = 1 - \exp(-2 \alpha(E) W),
\end{equation}
where $W$ is a thickness of the absorber, and $\alpha(E)$ is an absorption coefficient:
\begin{equation}
\label{eq:alpha}
    \alpha = \frac{4 \pi k}{\lambda}
\end{equation}
Here, we assume a perfect reflection at the bottom of the absorber doubling the length of the optical path, as reflected by the factor 2 in the exponential in Eq. (\ref{eq:absorptivity}).
The extinction coefficient $k$ is the imaginary part of a complex refractive index $\tilde{n}$:
%
\begin{equation}
\begin{split}
    &\tilde{n}   = n + i k,\\
    &\tilde{n}^2 = \tilde{\varepsilon}_\mathrm{r} = \varepsilon_1 + i \varepsilon_2,
\end{split}
\end{equation}
where $\tilde{\varepsilon}_\mathrm{r}$ is the complex dielectric function
whose real and imaginary components are $\varepsilon_1$ and $\varepsilon_2$, respectively.

As we assume that one absorbed photon generates one electron-hole pair following the SQ model,
the short-circuit current $J_\mathrm{SC}$ of a solar cell 
is given by the absorbed photon flux multiplied by an elementary charge $\mathrm{q}$:
\begin{equation}\label{eq:JSC}
    J_\mathrm{SC}\left( W \right) = \mathrm{q} \int_{0}^{\infty} %
    a \left( E; W \right) 
    \Phi_\mathrm{sun}\left( E \right) \, \dd E, 
\end{equation}
where $\Phi_\mathrm{sun}(E)$ is solar photon flux density at the photon energy $E$ .
From Eqs. (\ref{eq:absorptivity}) and (\ref{eq:JSC}),
it is evident that $J_\mathrm{SC}$ in Eq. (\ref{eq:absorptivity}) approaches the SQ limit
for an infinitely thick absorber ($ W \to \infty $).
This condition, however, cannot be met with the conventional first-principles electronic structure approaches where a coarse $k$-point mesh and an interpolation scheme, such as a Gaussian smearing function, are adopted to integrate the electronic Brillouin zone.
Due to the Gaussian tail below the bandgap,
a significant amount of photons with energy lower than the bandgap are absorbed, 
and $J_\mathrm{SC}$ approaches the SQ limit with a lower effective bandgap.
The correct SQ limit can be recovered with the aid of an abrupt absorption edge. 
To achieve this, we adopt a fine $k$-point mesh using a Wannier interpolation scheme.

%
Excited charge carriers can recombine via emitting photons, which is unavoidable
as radiative recombination is the time-reversal of light absorption. 
The radiative recombination rate $R_\mathrm{rad}$ for a solar cell at temperature $T$ is given by 
\begin{equation}
\begin{split}
    R_\mathrm{rad}(V) = & \frac{2\pi}{c^2 h^3} \int_{ 0 }^{\infty}
    a \left( E; W \right) \left[ \mathrm{e}^{\flatfrac{E - \mathrm{q}V}{k_\mathrm{B} T}}-1 \right]^{-1} 
    E^2 \dd{E} \\ \approx &
    \frac{2\pi}{c^2 h^3} \mathrm{e}^{\frac{\mathrm{q}V}{k_\mathrm{B} T}} \int_{ 0 }^{\infty}
    a \left( E;W \right) \left[ \mathrm{e}^{\flatfrac{E}{k_\mathrm{B} T}}-1 \right] ^{-1}
     E^2 \dd{E} \\ = &
     R_\mathrm{rad}(0) \mathrm{e}^{\frac{\mathrm{q}V}{k_\mathrm{B} T}} ,
\end{split}
\end{equation}
where $V$ is an applied voltage providing excess carriers.

The net radiative recombination is the difference between radiative recombination and the ambient irradiation 
whose rate is equal to the radiative recombination rate $R_\mathrm{rad}(0)$ at short-circuit conditions.
The net current density $J^\text{rad}$ limited by radiative recombination is thus given by
\begin{equation}\label{eq:jv_rad}
	J^\mathrm{rad}(V; W) = J_\mathrm{SC}(W) + J_{0}^\mathrm{rad}(W)(1 - \mathrm{e}^{\frac{\mathrm{q}V}{k_B T}}),
\end{equation}
where the saturation current $J_{0}^\text{rad} = \mathrm{q} R_\mathrm{rad}(0)$.

The dominant loss mechanism in a solar cell is often non-radiative recombination mediated by defects in the bulk or at interfaces.
The non-radiative recombination rate $R_\textrm{SRH}$ can be can be calculated using Shockley-Read-Hall statistics.\cite{Shockley:1952it,Hall:1952iz}
Recently, we proposed a self-consistent protocol to determine $R_\textrm{SRH}$ using the concentration, activation energy, and capture coefficient of point defects based on first-principles.\cite{Kim:2020gj}
Thus the current is further reduced from the radiative limit (Eq. \ref{eq:jv_rad}):
\begin{equation}\label{eq:jv}
\begin{split}
    J(V; W) =& J_\mathrm{SC}(W) + J_\mathrm{0}^\mathrm{rad}(W)(1 - \mathrm{e}^{\frac{\mathrm{q}V}{k_\mathrm{B} T}})\\
            &- q R_\mathrm{SRH}(V) W,
\end{split}
\end{equation}
where the maximum efficiency is given by:
\begin{equation}
	\eta_{max} = \text{max}_V \left(\frac{JV}
    {q \int_{0}^{\infty} E \Phi_\text{sun}\left( E \right)  \dd{E} } \right),
\end{equation}
which we call a trap-limited-conversion efficiency or an absorption-trap-limited conversion efficiency (aTLC) when finite absorptivity is considered.

\textit{Ab initio methods:}
Density functional theory (DFT) has become a workhorse technique in computational materials science that provides access to the total energy and electronic structure for interacting systems of ions and electrons. 
The bandgap of a solid (E$_g$) can be straightforwardly determined by mapping out the band structure along high-symmetry paths of the Brillouin zone and identifying critical points in the resulting $E(k)$ relations.

The optical absorption coefficient can in turn be determined from the frequency-dependent dielectric function (Eqn. \ref{eq:alpha}).
This function be calculated in several ways, the most convenient being a sum over dipole-allowed valence to conduction band transitions from Fermi's golden rule.\cite{gajdovs2006linear}
This approach works well for semiconductors without strong excitonic features that require an explicit description of electron-hole interactions. 

In this study, the total energy of pristine and defective crystals was calculated from DFT
using the projector-augmented wave (PAW) method \cite{Blochl1994,Kresse1999} and the hybrid exchange-correlation functional of Heyd-Scuseria-Ernzerhof (HSE06),\cite{Heyd2003} as implemented in \code{VASP}.\cite{Kresse1999}
We used fractions of screened exact exchange of 25.3\%, 26.8\%, and 27.8\% that reproduce the experimental bandgaps of \ce{Cu2ZnSnS4}, \ce{Cu2ZnSnSe4}, and \ce{Cu2ZnGeSe4}, respectively.
The wave functions were expanded in plane waves up to an energy cutoff of 380 eV.
A Monkhorst-Pack $k$-point mesh~\cite{Monkhorst1976} with a grid spacing less than $2\pi\times$0.03 \AA$^{-1}$ was used for Brillouin zone integration in the initial DFT calculations.

For defect formation, the same setup as previously reported was utilised.\cite{Kim:2020gj} 
A $2\times2\times2$ supercell expansion (64-atoms) of the conventional kesterite cell was employed. 
The all-electron wave functions were derived from the pseudo wave functions and atom-centered partial waves in the PAW method, and the overlap integrals were performed in real space using \code{pawpyseed} \cite{Bystrom:2019er} to calculate the electron-phonon coupling elemenets outlined by Alkauskas \textit{et al.}~\cite{Alkauskas:2014kk}.

For the high-frequency dielectric function, the energy eigenvalues and the matrix elements were calculated with fine \textit{k}-point mesh ($180\times180\times180$) in steps of 5 meV
using the Wannier interpolation approach implemented in \code{Wannier90}.\cite{Pizzi2020}
We have employed initial projections over Cu 3\textit{d}, Cu 4\textit{s}, Zn 4\textit{s}, Sn 5\textit{s}, Sn 5\textit{p}, Ge 5\textit{s}, Ge 5\textit{p}, and \ce{sp^3} orbitals of the chalcogens.
We verified that the interpolated band structures reproduce the original DFT results close to the band edges,
which ensures that all the optical transitions within the range of energy of interest (up to 5 eV) are well described by Wannier-function interpolation.

\begin{figure}[tb]
	\includegraphics[width=8cm]{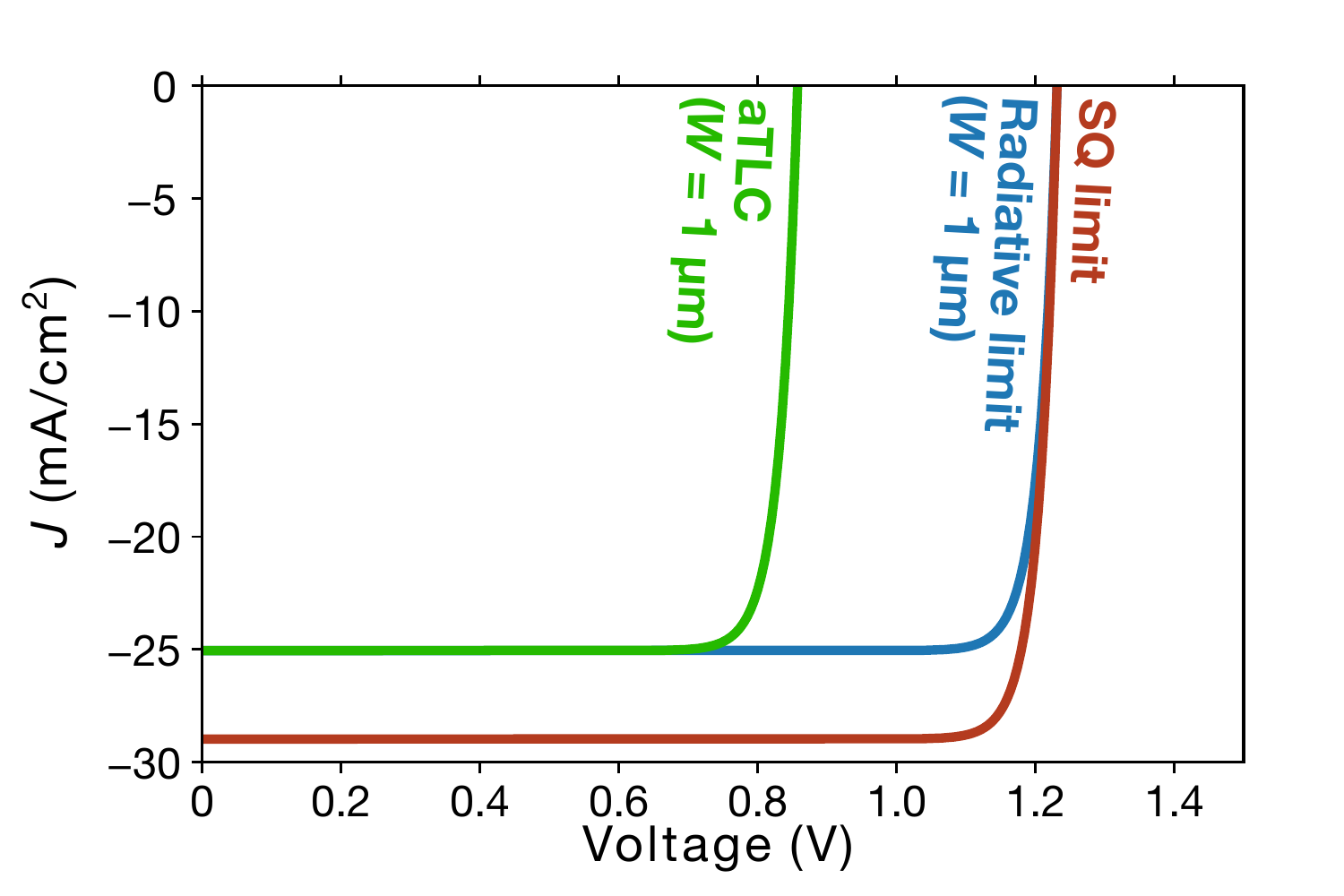}
	\caption{Current-voltage characteristics of CZTS
		calculated from the Shockley-Queisser limit (red line), radiative limit (blue line), and  aTLC (green line).
	The film thickness is assumed to be \SI{1}{\micro\meter} under AM1.5g illumination.
	}\label{fig:jv}
\end{figure}

\textit{Application to kesterites:}
The SQ efficiency limit of \ce{Cu2ZnSnS4} (CZTS) is as high as 32.1\%, using the AM1.5g solar spectrum, owing to the optimal bandgap of 1.5 eV.
An associated \JSC of \SI{28.9}{\mA\per\cm^2} is calculated.
Only radiative recombination is considered here, 
resulting in the high \VOC and fill factor of \SI{1.23}{\V} and 90\%, respectively (see Table \ref{tb:params} and Fig. \ref{fig:jv}).

%

The calculated optical absorption coefficient of CZTS
is shown in Fig. \ref{fig:alpha}.
Due to the finite absorptivity, the number of absorbed photons and the short-circuit current density decrease (see Fig. \ref{fig:jv}).
However, since CZTS is a direct bandgap semiconductor, the absorption coefficient is high with a sharp onset at the bandgap of \SI{1.5}{eV}.
$\alpha$ rises rapidly at the bandgap and exceeds $\SI{E4}{\per\cm}$,
which corresponds to the absorptivity $a=86.5\%$,
when the thickness of the absorber is \SI{1}{\micro\meter} (see Eq. \ref{eq:absorptivity}).
For \SI{2}{\micro\m} thickness, the absorptivity is greater than 98\%.
Thus \SI{2}{\micro\meter}-thick CZTS exhibits a radiative limit close to the SQ limit,
indicating the finite absorption does not limit the efficiency of standard CZTS solar cells, as expected.\cite{siebentritt2013kesterite}
The calculated absorption spectra agree well with previous calculations\cite{PhysRevMaterials.4.035402,PhysRevMaterials.2.085404} and experimental results\cite{Norman:2012jy,Hirate:2015ey}
except for the Urbach tail as our model does not include the disorder and defects.
Note that we do not consider weaker phonon-assisted absorption, which is important in indirect bandgap semiconductors such as Si.\cite{Noffsinger:2012hf,Zacharias:2015ex}

\begin{figure}[b]
	\includegraphics[width=8cm]{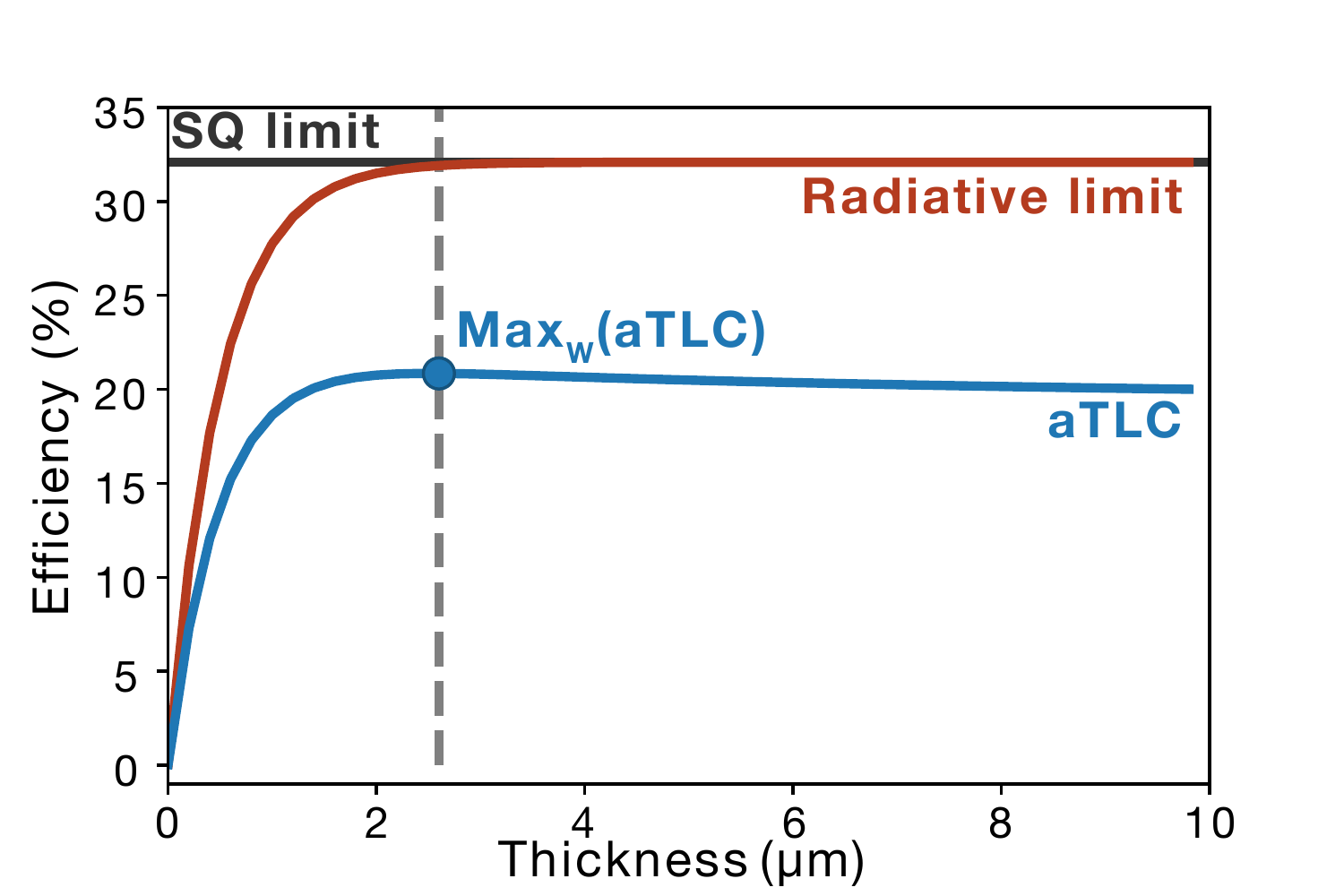}
	\caption{Thickness dependent maximum efficiencies based on the radiative limit and the aTLC measure of CZTS.
		The Shockley-Queisser limit is denoted by a black line.
		The optimal thickness that predicts the maximum aTLC is represented by a grey dashed line.
	}\label{fig:eff_w}
\end{figure}
%

Kesterite solar cells suffer from short carrier lifetime.
One contribution to the short carrier lifetime is native bulk defects causing rapid non-radiative recombination.
Our previous calculations showed that Sn reduction ({Sn(IV) towards Sn(II)}) is responsible for both deep levels and large carrier capture cross-sections.\cite{Kim:2018jd, Kim:2019iv, Li:2019bx}
Moreover, the narrow phase window of kesterites, mainly due to the Zn binary compound, 
offers little room for defect engineering reducing concentrations of the detrimental defects.

The most detrimental recombination centers have previously been identified as \ce{V_S}, \ce{Sn_{Zn}}, and their complexes. 
The defect concentrations can be tuned through the chemical environment.
The best condition is Cu-poor and Zn-rich growth with rich anion incorporation,\cite{Dimitrievska:2016cj}
which limits the formation of \ce{V_S} and \ce{Sn_{Zn}}.
The overall efficiency loss due to the non-radiative recombination was previously estimated to be about 10\% using TLC without consideration of finite absorptivity.

As we combine the non-radiative recombination and the radiative limit,
a significant \VOC loss is seen (Fig. \ref{fig:jv}).
The efficiency decreases by  9\% at the thickness of \SI{1}{\micro\meter}.
For thicker films, both the absorption and the non-radiative losses increase.
While the non-radiative recombination current increases linearly with the thickness (Eq.(\ref{eq:jv}),
the absorptivity increases rapidly and approaches to unity when the thickness is around \SI{2}{\micro\meter}.
Thus, in this approach, the maximum efficiency of 20.8\% is achieved at $W=\SI{2.6}{\micro\m}$ (see Table \ref{tb:params}).
We noted that the maximum aTLC values at their optimal thicknesses in Table \ref{tb:params} differ from corresponding TLC values\cite{Kim:2020gj} by around 1\% due to the aforementioned high absorption coefficient.
Further increase in the thickness causes efficiency loss due to enhanced non-radiative recombination.  

We note that aTLC still makes a number of strong assumptions that could be loosened to develop more quantitative models. 
There is an implicit assumption of infinite carrier mobility where the diffusion length does not limit carrier collection.
Taking into account finite reflectivity at the top and bottom surfaces, interface recombination, parasitic absorption effect in the window layer, and shunt and series resistance would further improve the predictive power of our approach. \cite{Guillemoles:2019gk}
These extensions would come at the cost of increased model complexity and the need to consider specific device architectures.

\begin{table}[]
	\centering
	\caption{Performances of kesterite solar cells predicted by SQ limit and aTLC (optimized thickness in parenthesis). The experimental results (Exp.) from the best reported solar cells are also listed. The conventional abbreviations are used where A=Ag, C=Cu, Z=Zn, T=Sn, G=Ge, respectively, e.g. AZTSe = \ce{Ag2ZnSnSe4}.
}\label{tb:params}
	\begin{tabular*}{\columnwidth}{lllllll}
		   \hline
		   \hline
           & $E_\mathrm{gap}$  & $\eta$  & \JSC  & \VOC       &  FF & Reference \\
           & (eV) & (\%) & ($\si{mA\per cm^{2}}$) & ($\si{V}$) & (\%) & \\
	   \hline
	   \hline
CZTS       & 1.50 & 32.1 & 28.9 & 1.23 & 90.0 & SQ limit \\
CZTSe      & 1.00 & 31.6 & 47.7 & 0.77 & 85.7 & SQ limit \\
CZGSe      & 1.36 & 33.3 & 34.3 & 1.10 & 89.1 & SQ limit \\
AZTSe      & 1.35 & 33.7 & 34.7 & 1.09 & 89.0 & SQ limit \\
	   \hline
CZTS       & 1.50 & 20.8 & 28.8 & 0.84 & 86.5 & aTLC (\SI{2.6}{\micro\meter}) \\
CZTSe      & 1.00 & 20.6 & 47.7 & 0.53 & 81.2 & aTLC (\SI{2.2}{\micro\meter}) \\
CZGSe      & 1.36 & 23.9 & 34.1 & 0.81 & 86.2 & aTLC (\SI{2.4}{\micro\meter}) \\
	   \hline
CZTS       & 1.50 & 11.0 & 21.7 & 0.73 & 69.27 & Exp.\cite{Yan:2018dw} \\
CZTSe      & 1.00  & 11.6  & 40.6   & 0.42  & 67.3  & Exp.\cite{Lee:2014cna} \\
CZTSSe     & 1.13 & 12.6 & 35.4 & 0.54 & 65.9  & Exp.\cite{Son:2019hm} \\
CZTGSe     & 1.11 & 12.3  & 32.3   & 0.53  & 72.7  & Exp.\cite{Kim:2016jr}  \\
CZGSe      & 1.36 & 7.6   & 22.8   & 0.56  & 60    & Exp.\cite{Choubrac:2018ex}  \\
AZTSe      & 1.35 & 5.2   & 21.0     & 0.50  & 48.7  & Exp.\cite{Gershon:2016kt}  \\
CZTS:Ag,Cd  & 1.40 & 10.1  & 23.4   & 0.65   & 66.2  & Exp.\cite{Hadke:2018hx} \\
	   \hline
	   \hline
	\end{tabular*}
\end{table}

In conclusion, we have presented
an \textit{ab initio} method for determining the maximum efficiency of a solar cell.
The absorption-and-trap-limited conversion efficiency (aTLC), 
procedure takes into account both radiative and non-radiative recombination.
Due to the high optical absorption coefficient of \ce{Cu2ZnSnS4},
the radiative limit approaches the SQ limit with a film thickness of \SI{2}{\micro\meter}.
The increase in non-radiative recombination current through point defects counterbalances this gain of absorption, 
which decreases the net efficiency of a thick film.
We determined the optimal thickness of an absorber layer of a solar cell based on its bulk properties.
As such, aTLC can be used a qualitative metric to evaluate and optimise the potential of emerging photovoltaic materials.

\textit{Data Availability Statement:}
The underlying models for capture cross-sections are implemented in the open-source package \code{CarrierCapture}.\cite{kim2020carriercapture}

\acknowledgements
We thank Jose A. Marquez and Thomas Unold for useful discussions and suggesting this extension of TLC.
Via our membership of the UK's HEC Materials Chemistry Consortium, which is funded by EPSRC (EP/L000202), this work used the ARCHER UK National Supercomputing Service (http://www.archer.ac.uk).
This research was also supported by the Creative Materials Discovery Program through the National Research Foundation of Korea (NRF) funded by Ministry of Science and ICT (2018M3D1A1058536).

%

\end{document}